\documentstyle[11pt]{article}
\def\be {\begin{eqnarray}}
\def\ee {\end{eqnarray}}
\def\beq {\begin{equation}}
\def\eeq {\end{equation}}
\def\bi {\begin{itemize}}
\def\ei {\end{itemize}}
\def\ben {\begin{enumerate}}
\def\een {\end{enumerate}}
\def\ni {\noindent}
\def\sutwo {${\rm SU(2)}_{\rm L} \times {\rm SU(2)}_{\rm R}$ }

\def\del {\partial}
\def\Tr {{\rm Tr}}
\def\om3 {{\cal O}(m_\pi^3)}
\newcommand{\nno}{\nonumber}

\newcommand{\should}{\stackrel{!}{=}}

\newcommand\half{{1\over 2}}

\newcommand{\equa}[1]{(\ref{#1})}

\newcommand{\Mpi}{m_\pi}

\newcommand{\Zeff}{Z_{\rm eff}}
\newcommand{\cO}[1]{{\cal O}({#1})}
\renewcommand{\thefootnote}{\fnsymbol{footnote}}
\begin{document}
\thispagestyle{empty}

\begin{center}
{\renewcommand{\arraystretch}{1.2}
{\Large\bf  IN-MEDIUM EFFECTIVE CHIRAL LAGRANGIANS\\[1mm]
AND THE PION MASS IN NUCLEAR MATTER\,\footnotemark[1]}\\[6mm]
{\large ANDREAS WIRZBA$^1$ and VESTEINN THORSSON\,$^2$}\\[5mm]
{\it $^1$  Institut f\"ur Kernphysik, Technische Hochschule Darmstadt,\\
Schlo{\ss}gartenstra{\ss}e 9, D--64289 Darmstadt, Germany}\\[0.5mm]
{\tt wirzba{\verb+@+}crunch.ikp.physik.th-darmstadt.de}
\\[1.5mm]
{\it $^2$ NORDITA, Blegdamsvej 17, DK--2100 Copenhagen \O, Denmark}\\[0.5mm]
{\tt thorsson{\verb+@+}nordita.dk}
}\end{center}

\vskip 1cm
\begin{minipage}[c]{13cm}{\renewcommand{\baselinestretch}{1.2}
\centerline{ABSTRACT}
\noindent
We argue that the effective pion mass in nuclear matter obtained from
chiral effective lagrangians is unique and does not depend on
off-mass-shell extensions of the pion fields as e.g.\ the PCAC
choice. The effective pion mass in isospin symmetric nuclear matter is
predicted to increase slightly with increasing nuclear density,
whereas the effective time-like pion decay constant and the magnitude
of the density-dependent quark condensate decrease appreciably. The
in-medium Gell-Mann-Oakes-Renner relation as well as other in-medium
identities are studied in addition. Finally, several constraints on
effective lagrangians for the description of the pion propagation in
isospin symmetric, isotropic and homogenous nuclear matter are
discussed.  }
\end{minipage}
\vfill
{\large
\begin{tabbing}
February 1995\` TH Darmstadt preprint IKDA 95/4 \\
\` {\normalsize {\tt hep-ph/9502314}}
\end{tabbing}}

\vskip 1cm

\footnotetext[1]{Talk  presented at
{\em Hirschegg '95: Hadrons in Nuclear Matter},
Hirschegg, Austria, January 1995}
\renewcommand{\baselinestretch}{1.0}
\renewcommand{\arraystretch}{1.0}
\setcounter{page}{0}
\setcounter{footnote}{0}
\newpage
\renewcommand{\thefootnote}{\#\arabic{footnote}}
\begin{center}
{\large\bf  IN-MEDIUM EFFECTIVE CHIRAL LAGRANGIANS
AND THE PION MASS IN NUCLEAR MATTER
}\\[5mm]
{ A. WIRZBA$^1$ and V. THORSSON\,$^2$}\\[4mm]
{\small\it $^1$  Institut f\"ur Kernphysik, Technische Hochschule Darmstadt,\\
Schlo{\ss}gartenstra{\ss}e 9, D--64289 Darmstadt, Germany\\[1mm]
$^2$ NORDITA, Blegdamsvej 17, DK--2100 Copenhagen \O, Denmark}
\end{center}

\vskip 0.6cm
\begin{minipage}[c]{13cm}\renewcommand{\baselinestretch}{1.0}
{\small \centerline{ABSTRACT}
\noindent
We argue that the effective pion mass in nuclear matter obtained from
chiral effective lagrangians is unique and does not depend on
off-mass-shell extensions of the pion fields as e.g.\ the PCAC
choice. The effective pion mass in isospin symmetric nuclear matter is
predicted to increase slightly with increasing nuclear density,
whereas the effective time-like pion decay constant and the magnitude
of the density-dependent quark condensate decrease appreciably. The
in-medium Gell-Mann-Oakes-Renner relation as well as other in-medium
identities are studied in addition. Finally, several constraints on
effective lagrangians for the description of the pion propagation in
isospin symmetric, isotropic and homogenous nuclear matter are
discussed.}
\end{minipage}

\vskip 0.8cm
\noindent{\large\bf 1.\  Introduction}\\[1mm]
The suggestion of Kaplan and Nelson\cite{kapnel} that attractive
S-wave interactions between kaons and nucleons could lower the
effective mass of kaons to the extent that kaons could condense in
dense neutron star matter at several times nuclear saturation density
has started in recent years a considerable discussion on the behaviour
of the Pseudo-Goldstone bosons of strong interaction physics, the
kaons and also pions, in dense nuclear matter.  Whereas the kaon case
is plagued by a lot of additional complications as the role of
resonances such as the $\Lambda(1405)$, which governs low-energy
$K^{-} p $ scattering, the coupling to the $\Sigma \pi$ channel and
the large size of the kinematical regions over which smoothness
assumptions are postulated to hold, the S-wave pion propagation in
symmetric nuclear matter is a much cleaner case~\cite{bkr,dee,blrt}
and can therefore serve as test ground for applying chiral
perturbation theory ideas to finite nuclear densities.

Recently, several authors have claimed that the method used to
motivate and describe meson condensation is incorrect\cite{dee,sc1}.
They argued that chiral effective lagrangians are inconsistent with
current algebra and PCAC\cite{dee,sc1}.  Secondly, they claimed that
the incorporation of these off-meson-mass-shell amplitudes in the
calculation inevitably leads to an effective repulsion which serves to
inhibit meson condensation\cite{sc1}.

In a previous letter \cite{we} we showed that these claims do not
hold, as either incomplete chiral lagrangians were considered or the
source coupling was done inconsistently.  In fact, in Ref.\cite{we} it
was shown -- in line with well-established theorems\cite{ccwz} -- that
the S-wave meson-nucleon scattering amplitudes obtained
off-meson-mass-shell are entirely unphysical as they are subject to
the choice of the meson field, and are thus not to be viewed as
constraints on a theory.  Furthermore, it was shown that the effective
meson mass in nuclear matter is {\it independent} of the choice made
for the meson field.
This is to be understood as a consequence of a general
rule:  any physically relevant observable is independent of
the choice of meson field variables, as is the case for S-matrix
elements\cite{ccwz}.
This conjecture was supported by two calculations \cite{we}: one
using a formulation of chiral perturbation theory
for which the canonical meson field {\it is} to be identified with
the divergence of the axial vector current, namely that originating
in the work of Gasser and Leutwyler\cite{gl},
and one using the traditional treatment, originally
due to Kaplan and Nelson\cite{kapnel}, in which the meson
field {\it is not} to
be identified with the divergence of the axial vector current.

Below, we discuss how chiral perturbation theory can be applied to the
analysis of S-wave pion propagation.  We consider tree level
lagrangians throughout, working to ${\cal O}(Q^2)$. In section 3 we
illustrate our results for homogeneous, isotropic, isospin symmetric
and spin-unpolarized nuclear matter, and evaluate nucleon operators in
the mean field approximation, such that the corresponding results hold
modulo nuclear correlation corrections.  We work out the in-medium
pion mass, the effective pion decay constant, the in-medium quark
condensate, the Gell-Mann-Oakes-Renner relation and the PCAC relation
in nuclear matter.  In section 4 we discuss how the new developments
about non-relativistic chiral lagrangians \cite{leutnonrel} and
generalizations to four-quark condensates \cite{ks} can constrain the
structure of the in-medium chiral lagrangians.

\vskip 0.5cm
\noindent{\large\bf 2.\  Chiral Perturbation Theory and S-wave Pion
Propagation}\\[1mm]
%
Here, we briefly review the functional integral formulation of chiral
perturbation theory developed by Gasser and Leutwyler\cite{gl}, which
was extended to include nucleons by Gasser, Sainio and
\u{S}varc\cite{gss}.  In this approach, the effective generating
functional for QCD Green functions, the vacuum-to-vacuum transition
amplitude $\exp(i Z_{\rm QCD })$, is developed as follows.  External
color-neutral sources, the isovector vector $v_\mu$, the isovector
axialvector $a_\mu$, the isoscalar scalar $s$ and isovector
pseudoscalar $p$, are coupled to the corresponding quark currents of
the QCD action and assigned chiral \sutwo transformations such that
the source-extended action is {\it locally} chiral invariant.  Note
the current quark masses of the QCD lagrangian are hidden in the
scalar isoscalar coupling $\bar q s q$ by $s$ containing a constant
piece: $s=s_0+s'$ where $s_0$ is the quark mass matrix $\cal M$.  The
fact that the sources are coupled in this way ensures that chiral
QCD-Ward identities are satisfied.  The central idea is now that the
generating functional for the low-energy effective theory on the
hadronic level, $\exp(i Z_{\rm eff})$, should depend on the {\em
same\/} external sources.  In this way, one identifies the
to-be-determined Ward identities of QCD with those of the low-energy
effective theory. As the effective theory is a that stage -- and {\em
has to be} -- completely general (besides its local chiral
invariance), further empirical facts are needed in order to constrain
the effective generating functional to something useful and workable.
Naturally, the effective theory should respect the usual invariances
under e.g.\ Lorentz-transformations, parity- and time-reflection,
charge-conjugation and should be also local. Its sources and fields
have to be color-neutral. All the hadronic fields entering the
effective action are dummy fields as they are integrated over in the
generating functional formalism. At low energies they can be limited
to the pions (and kaons), as the masses of these Pseudo-Goldstone
bosons are much smaller than the masses of all the other hadrons
(which are of the order of the chiral symmetry breaking scale $\Lambda
\simeq 1$~GeV) and as therefore the low-momentum decoupling theorems
for Goldstone bosons hold: at low $Q$ the Goldstone bosons are weakly
interacting, the hadronic Green's functions are dominated by poles due
to Pseudo-Goldstone boson exchange, and the vertices in the
Goldstone-Goldstone interaction admit a Taylor series expansion in
powers of the small momenta $Q$.  In Weinberg's chiral counting
scheme\cite{weinberg} the various tree and loop terms are ordered in
powers of the external momenta $Q$ and the pion mass $m_\pi$ (or kaon
mass $m_K$). The values of tree-level coefficients follow from
empirical input. The final ingredient in standard chiral perturbation
theory is the conjecture that the non-vanishing quark condensate
$\langle 0| \bar q q|0\rangle$ is not only the order parameter of
spontaneous breaking of chiral symmetry, but that it is so large that
higher quark condensates can be neglected at leading order~\cite{ks}.
This implies (see the discussion in section~4) that the current quark
mass matrix ${\cal M}$ scales as the squared pion mass and that the
Weinberg counting is in squared powers of the external momenta, $Q^2$,
and in linear powers of the average current quark mass, $\hat
m=(m_u+m_d)/2$.

At low-energies the hadronic action, $S_{\rm eff}$, is therefore
formulated in terms of a 2$\times$2 dummy field $U$ (which
transforms linearly under \sutwo), involving pions.  Note that in this
formalism, the sources $s$, $p$, $v_\mu$, and $a_\mu$ are coupled from
the start to quark bilinears and are therefore directly associated
with
\sutwo currents in quark variables.
The PCAC prescription thus emerges naturally as the sources are
transcribed to the hadronic level.  Similarly, one may introduce
nucleons, $N$($\bar N$), coupled to corresponding external sources,
$\bar\eta$ ($\eta$), which transform non-linearly under chiral
transformations in order to ensure that the coupling terms are
(locally) chiral invariant. Note in this case the caveat that the
sources $\eta$ and $\bar \eta$ are only defined on the hadronic level
as there does not exist any simple interpretation on the QCD level of
a baryonic source.  Furthermore, the generating functional $\exp(i
Z_{\rm eff})$ should not be linked to the vacuum-to-vacuum transition
amplitude any longer, but rather to the $n$-nucleon to $n$-nucleon
transition amplitude (with \mbox{$n\!\geq \!1$})~\cite{gss}.  It reads
\beq
e^{ {i}  Z_{\rm{eff}}[s,p,v_\mu,a_\mu,\eta,{\bar \eta}]}
= {\cal N}
\int {d} U \, {d}N \, {d}{\bar N}
{e}^{{i}  \int\! {d}^4x\, ( {\cal L}_{\pi \pi} + {\cal L}_{\pi N} +
{\bar \eta}N + {\bar N}\eta ) }
\,\,\,,
\label{zeff}
\eeq
\ni where $\cal N$ is an overall normalization, $U,N,{\bar N}$
are dummy fields, and the lagrangians $ {\cal L}_{\pi \pi} $ and $
{\cal L}_{\pi N} $ are dependent on the sources $s$, $p$, $v_\mu$ and
$a_\mu$.  In what follows, we shall use the scalar source $s$ to
generate the quark mass matrix, $s={\cal M}={\rm diag}(m_u,m_d)$, and
furthermore retain only the source for asymptotic pions, $p$, but set
$v_\mu=a_\mu=0$ (unless otherwise specified).  The sources $\bar \eta$
and $\eta$ generate one-nucleon in- and out-states.  The nucleons are
treated in the static fermion formalism\cite{jenman}, in which nucleon
loops play no role ( see also appendix A of Ref.\cite{bkkm} ), and the
nucleon determinant may therefore taken to be unity.

The lagrangian entering in the generating functional (\ref{zeff}),
${\cal L}_{\rm{eff}}={\cal L}_{\pi \pi} + {\cal L}_{\pi N}$,
is to leading order given by the nucleon kinetic energy term,
${i}\bar N({\rm v}\cdot
\del ) N$
and to subleading order, ${\cal O}(Q^2)$, by\cite{gss,bkm}
\be
{\cal L}_{\pi \pi}^{(2)} &=&
\frac{f_\pi^2}{4} {\rm Tr} \del_\mu U \del^\mu U^\dagger
+ \frac{f_\pi^2}{4} {\rm Tr} ( U^\dagger \chi\mbox{+}\chi^\dagger U )
\,\,\,,
\label{lpp2} \\
{\cal L}_{\pi N}^{(2)} &=&
- \frac{\sigma}{4m_\pi^2} {\bar N}N {\rm Tr}
( U^\dagger \chi\mbox{+}\chi^\dagger U )
+ c_2 {\bar N}({\rm v} \cdot u)^2 N + c_3 {\bar N} (u \cdot u) N
\,\,\,,
\label{lpn2}
\ee
where $u_\mu=i u^\dagger \del_\mu U u^\dagger$, $U=u^2=\exp(i \tau^a
\pi^a/f_\pi)$, $\chi=2 B (s+i p)$, and ${\rm v}_\mu$ is the
four-velocity of the nucleon which reduces to ${\rm
v}_\mu$=$(1,0,0,0)$ in the rest frame of the nucleon. The empirical
values\footnote{The pion decay constant $f_\pi$= 93 MeV and the pion
mass $m_\pi$ = 139 MeV.} of $f_\pi^2$ (and the later to be used
$m_\pi^2$) include $\cO{Q^2}$ corrections to the corresponding
quantities at tree level. They are used here for notational
convenience as we anyhow neglect corrections to the lagrangian of
$\cO{Q^3}$ and higher.  The constants $\sigma$, $c_2$ and $c_3$ are
linear in the quark masses and therefore of order $\cO{Q^2}$. The
constant $\sigma$ is to be identified with the sigma term,
$\sigma$($t$=0), which also serves to increase the nucleon mass over
that in the SU(2) chiral limit, $m_N$=$m_0\mbox{+}\sigma$, where $m_0$
$\approx$ 890~MeV, using $\sigma$ = 45~MeV~\cite{gls}. Thus the sigma
term is fixed to be positive.  We do not write down the Weinberg
(vector) term explicitly in the lagrangian above, as it does not enter
in the S-wave pion propagation in isospin-symmetric matter to be
discussed in the next section.

Expanding $U$ to second order in $\pi^a$, we find
\be
{\cal L}_{\rm{eff}} &=& \bar N (i {\rm v} \cdot \del -\sigma) N
+\half (\del_\mu \pi)^2 - \half m_\pi^2 \pi^2 \nno \\
&& \mbox{} + \frac{1}{f_\pi^2}\left (\half\sigma
 \pi^2\mbox{+}c_2({\rm  v}\cdot \del \pi)^2\mbox{+}c_3
(\del_\mu \pi)^2 \right )
 \bar N N   + j^a \pi^a
\left( 1\mbox{$-$}\frac{\sigma {\bar N}N}{f_\pi^2 m_\pi^2}\right )
\,\,\,.
\label{lgl}
\ee
The pseudoscalar source is $j^a \equiv 2 B f_\pi p^a$, in terms of the
original source $p^a$ ($p$=$p^a \tau^a$), and of the ``quark
condensate'' $-2 f_\pi^2 B$=$-2 f_\pi^2 m_\pi^2/(m_u+m_d)$, as follows
from the quadratic expansion.  Since Green functions are obtained by
taking functional derivatives of the generating functional with
respect to the source $j^a$, the nontrivial coupling of the source to
the pion field in Eq.(\ref{lgl}) plays an important role in the
consistent description of the pion-off-shell S-wave $\pi N$ amplitude,
see \cite{we}. From the lagrangian \equa{lgl} we find the isospin even
scattering length, $a^{+}_{\pi N}$:
\beq
 a^{+}_{\pi N} = \left \{4\pi f_\pi^2(1+m_\pi/m_N) \right \}^{-1}\left
(2(c_2+c_3)m_\pi^2+\sigma\right) + \cO{m_\pi^3} \ .
 \label{scattlength}
\eeq
Empirically, $a_{\pi N}^{+}= -0.0083 m_\pi^{-1}$ \cite{koch},
corresponding to a repulsive interaction. Using $\sigma$ $\approx$ 45~MeV, we
find $(c_2\mbox{+}c_3)m_\pi^2$ $\approx$ $-$26~MeV. Improved values for the
constants can be found by including loop corrections \cite{bkm}: The
first corrections of $\cO{m_\pi^3}$ result from finite loop
terms. This is the reason why in nuclear matter all quantities of
order $\cO{Q^2}$ get their first correction already at $\cO{Q^3}$ and
not at $\cO{Q^4}$ as their free-space analogs.

\vskip 0.5cm
\noindent{\large\bf 3.\ The Effective Meson Mass in Nuclear Matter}\\[1mm]
Given the problems encountered in extending chiral perturbation theory
from the meson to the baryon sector, it is not surprising that a
rigorous formulation of the expansion in nuclear matter has not yet
been found.  The presence of an additional scale (the Fermi momentum
of nucleons), the breaking of Lorentz-invariance, and nuclear
correlations, add new levels of complexity to the formulation of a
chiral expansion.  As a first step, one simply uses a free space
chiral expansion, such as those outlined above and evaluates nucleon
operators at the mean field level, and consequently works with the
action to linear order in density: namely, $S=\int d^4 x {\cal
L}^{(2)}(\rho)$ with
\beq
{\cal L}^{(2)}(\rho)  =
	  \frac{f_\pi^2}{4}\left(g^{\mu\nu}
       \mbox{+}\frac{D^{\mu\nu}\rho}{f_\pi^2} \right )\! {\rm Tr}
          (\del_\mu U \del_\nu U^\dagger )
       + \frac{f_\pi^2}{4} \left ( 1\! -\!
 \frac{\sigma\rho}{f_\pi^2m_\pi^2}\right )\!
          {\rm Tr}(U^\dagger \chi\mbox{+}\chi^\dagger U),
\label{lrho}
\eeq
and $D^{\mu \nu}\! \equiv\! 2 c_2 {\rm v}^\mu {\rm v}^\nu\mbox{+}2 c_3
g^{\mu \nu}$, as follows from Eqs.(\ref{lpp2}) and (\ref{lpn2}).  One
may raise the question of the role of the pion interpolating field in
this context.  We will argue that the basic idea, established
rigorously in the case of free space scattering \cite{ccwz}, that
physically relevant observables are independent of the choice of field
variables, also holds in nuclear matter. In this case, the relevant
observable is the position of the pole of the pion propagator in
symmetric nuclear matter\cite{polerefs}.  The pole position is often
referred to as the effective mass, which we shall also do here.

As alluded to above, in the nucleon mean field approximation we set
$\langle{\bar N}N\rangle=\rho$ (we also approximate the vector density
by the scalar density).  Note that from (\ref{zeff}) it is immediately
apparent that since $U$, and therefore $\pi$, is a dummy variable, any
observable must be independent of this field as long as it is
correctly normalized and has a nonvanishing matrix element between the
pion and the vacuum.  At tree level we have \footnote{$Z[j;\rho]$ :=
$Z_{\rm{eff}}^{\rm m.f.a.} [{\cal M},j,0,0,0,0]$ and the so-called
classical pion field: $\phi_\pi^a$ = $(1\!-\!\sigma\rho/f_\pi^2
m_\pi^2)\pi^a$.}  $Z[j;\rho]$=$S[\pi]\mbox{+} \int\! {d}^4x\,
j^a(x)\pi^a(x) (1-\sigma \rho/f_\pi^2 m_\pi^2)$. Thus the tree-level
effective action, $\Gamma[\phi_\pi;\rho]$=$Z[\phi_\pi;\rho]\mbox{$-$}
\int\! {d}^4x\, j \phi_\pi =
S[(1\mbox{$-$}\sigma\rho/{f_\pi^2 m_\pi^2})^{-1}\phi_\pi]$,
reads
\be
\Gamma[\phi_\pi;\rho] = \half \int {d}^4x\,
\left( 1\mbox{$-$}\frac{\sigma \rho}{ f_\pi^2 m_\pi^2 }
\right)^{\!\!-2}\!\!
\left(
\del_\mu \phi_\pi \del_\nu \phi_\pi
\left(\! g^{\mu \nu}\mbox{+}\frac{D^{\mu \nu} \rho }{f_\pi^2}\! \right)
\!-\! m_\pi^2
\left( 1\!
-\! \frac{\sigma \rho}{ f_\pi^2 m_\pi^2 }\! \right)\! \phi_\pi^2
\right)\, , \nonumber
\ee
\ni from which we obtain the in-medium charged pion propagator
\beq
D(q,\rho) =
\frac{ i \left( 1
- \frac{\sigma \rho}{ f_\pi^2 m_\pi^2 } \right)^2 }
{ q^2 - m_\pi^2 + \frac{\rho}{f_\pi^2}
( \sigma + 2 c_2 ({\rm v} \cdot q)^2 + 2 c_3 q^2 ) } +\om3 \ .
\label{gssprop}
\eeq
Evaluating the poles of the propagator we find the effective pion mass
${m_\pi^\ast}^2(\rho):=\mbox{$\omega^2({\bf k}\mbox{=}0;\rho)$}$ in
symmetric nuclear matter to be
\be
{m_\pi^\ast}^2(\rho)
&=& m_\pi^2 \left\{1- \sigma\,\rho/ (f_\pi^2 m_\pi^2) \right\}/\left(
                       {1+ 2(c_2+c_3)\,\rho /f_\pi^2}\right)
  +\om3
\label{mpifull} \\
       &=& m_\pi^2\left( 1 -\rho (
2(c_2+c_3)m_\pi^2
+\sigma) / (f_\pi^2 m_\pi^2) \right )
           + {\cal O}(m_\pi^3;\rho^2)
\,\,\,,
\label{mpilin}
\ee
\ni where in Eq.~(\ref{mpilin}) we explicitly show
the prediction to linear order in density.  Eq.~(\ref{mpilin}) gives
$m_\pi^\ast(\rho) = m_\pi - 2 \pi a^+_{\pi N} \rho / m_R$, to linear
order in density, where $m_R$ is the reduced mass of the pion-nucleon
system.  The behaviour of the effective mass to linear order in
density is as expected from the lowest order optical potential,
without reference to chiral perturbation theory.  The effective mass
receives additional contributions over those given by
Eq.(\ref{mpifull}) at higher than linear order in density, from
factors such as those mentioned at the beginning of this
section~\footnote{In the equations above, and in those which follow,
we exhibit explicitly only the density dependence following from the
chiral lagrangian, to ${\cal O}(Q^2)$, and in the mean-field
approximation.}.

The fact that the lagrangian
${\cal L}_{\rm{eff}}$
embodies the PCAC choice of the interpolating pion field  appears
in the residue of the pion propagator,  not in the effective
pion mass. A different off-shell choice for the pion field  implies a
change in the source coupling in \equa{lgl}  (~e.g. $j^a\pi^a$
\cite{kapnel} instead of
$j^a\pi^a(1-\sigma \rho/f_\pi^2 m_\pi^2)$~) and results to a different
residuum (~e.g. 1 instead of $(1-\sigma \rho/f_\pi^2 m_\pi^2)^2$~) of
the propagator \equa{gssprop}, but leaves the pole position invariant.
We therefore conclude that there is no discrepancy in the effective
mass obtained from theories using different interpolating pion
fields\cite{we}.  It is furthermore to be noted that, within the
approximations that we are working with and have stated above, the
effective mass \equa{mpifull} predicted by models with different
interpolating pion fields is identical {\it to all orders in nuclear
density}.

%
%

To conclude this section, we discuss the relation of the effective
pion mass in symmetric nuclear matter, to the Gell-Mann-Oakes-Renner
(GMOR) relation\cite{gmor}.  The GMOR relation reads $ f_\pi^2
m_\pi^2$ = $ -\frac{m_u+m_d}{2}
\langle 0 | \bar{u}u\mbox{+}\bar{d}d | 0 \rangle
+{\cal O}(m_\pi^4)$.
It is
derived from the effective second-order lagrangian
${\cal L}^{(2)}_{\pi \pi}$, Eq.(\ref{lpp2}), by the identification
$Z_{\rm{QCD}}\should Z_{\rm{eff}}$, which leads to
\beq
 \left. \frac{\delta Z_{\rm{QCD}} }{\delta {\cal M}(x)}
  \right|_{{\cal M}=0}
  = - \langle 0 | \bar u u + \bar d d | 0 \rangle = 2 f_\pi^2 B\ .
 \label{findgmor}
\eeq
Solving for $B$ and using the relation $B (m_u\mbox{+}m_d)\! =\!
m_\pi^2$ one thus obtains the GMOR relation\cite{gl}.  To study the
GMOR relation with respect to the in-medium pion mass, we use the
lagrangian ${\cal L}_{\rm{eff}}$ in the nucleon mean field
approximation, Eq.(\ref{lrho}).  Since matter breaks
Lorentz-invariance (but still keeps rotational invariance, if it is
isotropic), it is convenient to separate space and time components
\cite{kirch} via $f_\pi^2
\left(g^{\mu\nu}\mbox{+}{D^{\mu\nu}\rho}/{f_\pi^2} \right ) =
{f_t^\star}^2(\rho)g^{00}\mbox{+}{f_s^\star}^2(\rho) g^{ii}$ where the
time-component is given by
\beq
 {f_t^\star}^2 (\rho) = f_\pi^2
 \left(1 + { D^{00} \rho}/{f_\pi^2}\right )
  + {\cal O}(m_\pi)  \ .
\label{ftrho}
\eeq
Starting from Eq.(\ref{lrho}) and using the same method as to derive
Eq.(\ref{findgmor}), we obtain the density-dependent quark condensate
\beq
\langle \bar u u + \bar d d \rangle_\rho =
\langle 0 | \bar u u + \bar d d | 0 \rangle
\left( 1 - {\sigma\rho}/({f_\pi^2m_\pi^2})\right )
 +{\cal O}(m_\pi)
\,\,\,,
\label{qqrho}
\eeq
a result which is, in fact, model-independent\cite{birse}.
Eq.(\ref{qqrho}), when combined with the effective pion mass as
given in Eq.(\ref{mpifull}),
yields the in-medium GMOR relation in
the nucleon mean-field approximation:
\beq
{f_t^\star}^2(\rho)\, {m_\pi^\star}^2(\rho)
= - {\scriptstyle\frac{m_u+m_d}{2}}
\langle \bar{u}u+\bar{d}d \rangle_\rho
+ \om3 \,\,\,.
\label{gmorrho}
\eeq
(Other discussions of the GMOR relation at finite density are given in
Ref.\cite{gmorrho}.)\ It is therefore only the time-component of the
coupling constant, $f_t^\star$, that enters in the GMOR relation at
finite density.  As a function of density, $f_t^\star$ and $-\langle
\bar{u}u+\bar{d}d \rangle_\rho$ decrease appreciably (to about
two-thirds of their vacuum value at nuclear matter density), whereas
$m_\pi^\star$ increases, though very slowly.

It is worthwhile to check that $f_t^\star(\rho)$, as given by
Eq.(\ref{ftrho}), agrees with the definition in terms of the axial
current coupling to the pion in matter,
\beq
\langle 0 | \bar q \gamma_0 \gamma_5 (1/2) \tau^a q
| \pi^b\rangle_\rho = i p_0 \delta^{ab} f_t^\star(\rho)
+ \om3
\,\,\,.
\label{ftgen}
\eeq
As the pion state in the medium, the ket $| \pi^b\rangle_\rho$, is not
known, the expectation value is evaluated from the axial vector
two-point function where the information about the pion state doesn't
enter because of closure. The axial vector correlator reads at zero
nuclear density (see Ref.\cite{gl}):
\be
\left. \frac{\delta^2  \Zeff}
{\delta a_\mu^a(-q)\,  \delta a_\nu^b(q)}
\right|_{a=v=p=0;s={\cal M}} &=&
 i   \int d x\,
e^{i  q (x-y)} \langle 0| T A_\mu^a (x) A_\nu^b(y) |
0\rangle \nno \\
 &=& \delta^{ab}\left \{ g_{\mu\nu} f_\pi ^2
+ \frac{q_\mu q_\nu f_\pi^2}
{\Mpi^2 -q^2}
\right \} + {\cal O}(q^2) \,\,\,. \label{aa}
\ee \vskip -1mm \noindent
Eq.(\ref{aa}), and other correlators, may be evaluated at finite
density, in the mean field approximation, by re-instating the general
sources $s$, $v^\mu$ and $a^\mu$ in $Z_{\rm{eff}}$, expanding the
action to second order in the pionic field, and integrating out the
pionic degrees of freedom.  The second order variation with respect to
the external sources then gives the two-point function.  For example,
one finds that -- up to ${\cal O}(m_\pi)$ corrections -- the variation
with respect to $a_0$ gives an in-medium correlator of the same form
as the time-time-component of Eq.(\ref{aa}), but with $f_\pi^2$
replaced by ${f_t^\ast}^2 (\rho)$, Eq.(\ref{ftrho}), and $m_\pi$
replaced by the effective mass (\ref{mpifull}).  Thus the desired
equivalence is established. This result is independent of the
off-shell extension of the pion field, as the pseudoscalar sources,
$p^a$, do not enter in this calculation.

In an analogous manner one may arrive at other relations valid at
finite density.  Evaluation of the in-medium pseudoscalar correlator
results in
\beq
 {g_\pi^\star}^2 (\rho) = (2 B f_\pi)^2
    { \left (1 -{\sigma \rho}/{f_\pi ^2 \Mpi^2} \right )^2 }
                  / \left(  {1 +{D^{00} \rho}/{f_\pi^2}} \right )
  + {\cal O}(m_\pi)  \ ,
\label{gpirho}
\eeq
(~where $g_\pi$ in the vacuum is defined as
$g_\pi \delta^{ab}\! \equiv\!\langle 0|\bar q i\gamma^5 \tau^a q |
\pi^b\rangle $~).
This result is dependent on the off-shell extension of the pion field,
as the calculation explicitly involves functional derivatives with
respect to the pseudoscalar source, $p^a$. The
axialvector-pseudoscalar correlator is $p^a$-dependent, as well.
Using Eqs.(\ref{qqrho}) and (\ref{mpifull}) one can then check that
the finite-density version of the PCAC relation~\cite{gl} holds under
the source-coupling of \equa{lgl}:
\beq
  f_t^\star(\rho) {m_\pi^\star}^2(\rho)
 = {\scriptstyle \frac{m_u+m_d}{2} }\,g_\pi^\star(\rho)
  + \om3 \ .
\eeq

\vskip 0.5cm
\noindent{\large\bf 4.\ Non-relativistic Chiral Lagrangians}\\[1mm]
Note the statements of the last section about the S-wave pion
propagation in homogeneous, isotropic nuclear matter result from an
$\cO{Q^2}$ chiral lagrangian to linear order in the density in the
mean-field approximation.  Already at this order the presence of the
nuclear matter background distinguishes a preferred rest frame for the
pion propagation and therefore Lorentz-invariance is broken.  This
manifests itself e.g.\ in distinct values of the time-like and
space-like effective pion decay constant, $f_t^\ast(\rho)$ and
$f_s^\ast(\rho)$, respectively.  Thus in trying to build up a chiral
perturbation theory for nuclear matter -- valid for even higher
densities -- one cannot any longer insist on Lorentz-invariance as one
of the preconditions.  Rather it is has to be replaced by the
left-over three-dimensional Euclidean rotational invariance, in case
the background matter is still isotropic and homogenous. The question
is whether there exists a modified chiral perturbation theory under
such non-relativistic preconditions. Fortunately, in a different
context, namely solid state physics, such non-relativistic chiral
perturbation theory is already known, see Leutwyler's work
\cite{leutnonrel} and references therein.  In Ref.\cite{leutnonrel} an
effective field theory is constructed which is relevant for the low
energy analysis of spontaneously broken symmetries in the
non-relativistic domain and which applies to any system for which
Goldstone modes are the only excitations without energy gap. Let us
assume that the corresponding hamiltonian is symmetric with respect to
a Lie group $G$, whereas the ground state is only invariant under
the subgroup $H\subset G$. Therefore the effective theory involves
${\rm dim} G - {\rm dim} H$ real fields which we will still refer to
as `pion' fields, $\pi^a$, where small Latin indices $a=1,\dots,{\rm
dim} G - {\rm dim} H$ denote the components of the effective field.
As usual, external fields, the time-like $f_0^A$ and the space-like
$f_i^A$, are coupled to the charge density $J^0_A(x)$ and charge
current $J^i_A(x)$, respectively, where capital Latin indices
$A=1,\dots, {\rm dim } G$ label here the group generators.  The
effective lagrangian extended in this way by external sources is
expanded according to the low-energy scheme\cite{leutnonrel}
\be
 {\cal L}_{\rm eff} =
 \sum_{n_t,n_s} {\cal L}_{\rm eff}^{(n_t, n_s)} \ ,
 \label{lnonrel}
\ee
where the lagrangian ${\cal L}_{\rm eff}^{(n_t,n_s)}$ is of
$\cO{\omega^{n_t}, |{\bf q}|^{n_s}}$ (with $n_t$ and $n_s$ positive
integer), is chiral symmetric and rotational invariant.  The last
point implies that $n_s$ has to be even. The decoupling of the
Goldstone bosons at low energies and momenta excludes the effective
lagrangian ${\cal L}_{\rm eff}^{(0,0)}$ (which contains only pion
fields, but no derivatives).  The leading order effective lagrangian
(without explicit symmetry breaking) is therefore of $\cO{\omega,|{\bf
q}|^0}$ and collects only pion vertices with one time derivative as
well as those with one insertion of the time-like external source
$f^A_0$ which is counted as $\cO{\omega^1}$.  It
reads~\cite{leutnonrel} ${\cal L}_{\rm eff}^{(1,0)}$ = $c_a (\pi)
{\dot \pi}^a + e_A (\pi) f_0^A $.  The quantities $c_a(\pi)$ and
$e_A(\pi)$ are implicitly density- or background-dependent of $\cO{1}$
and make the lagrangian chirally invariant. The term ${\cal L}_{\rm
eff}^{(1,0)}$ cannot occur in Lorentz-invariant effective theories.
The next corrections are of $\cO{\omega^2,|{\bf q}|^0}$ and
$\cO{\omega^0, |{\bf q}|^2}$, respectively \cite{leutnonrel}:
\be
      {\cal L}_{\rm eff}^{(2,0)} &=&
         \half {\bar g}_{ab}(\pi) {\dot \pi}^a {\dot \pi}^b
        + {\bar h}_{a A}(\pi) f_0^A  {\dot \pi}^a
        +\half {\bar k}_{AB} (\pi) f_0^A f_0^B \label{ln20}\\
      {\cal L}_{\rm eff}^{(0,2)} &=&
         -\half { g}_{ab}(\pi) {\del_i \pi}^a {\del_i \pi}^b
        - {h}_{a A}(\pi) f_i^A  {\del_i \pi}^a
        -\half {k}_{AB} (\pi) f_i^A f_i^B\ , \label{ln02}
\ee
where $f_i^A$ is a space-like source of $\cO{|\bf q|^1}$ and ${\bar
g}_{ab}(\pi)$, ${g}_{ab}(\pi)$, ${\bar k}_{AB}(\pi)$, ${k}_{AB}(\pi)$ etc.\
are `time-like' and `space-like' metric tensors in the coset space and
group space, respectively. Lorentz-invariance would imply the
coincidence of the barred with the unbarred terms ($c\!\equiv\! 1$).
Note that $e_A(\pi)$ at $\pi\!=\!0$ gives a term in the effective
lagrangian which is linear in the external source and hence determines
the expectation value of the charge density in the ground state,
$e_A(0)= \langle gs| J^0_A(x) | gs\rangle$. For non-abelian symmetries
the charge densities transform non-trivially under the group $G$ such
that their expectation values can serve as order parameters. If the
charge density acquires a non-zero expectation value, as e.g.\ the
spin-density in the ferromagnet, the first order effective lagrangian
is non-vanishing and the dispersion follows the usual non-relativistic
law, $\omega \propto {\bf q}^2$.  If, however, $e_A(0)=0$ and
therefore the charge density does not acquire a non-zero expectation
value, as e.g.\ for an antiferromagnet, the effective lagrangian
${\cal L}_{\rm eff}^{(1,0)}$ vanishes identically (because of the
equation of motion and chiral Ward identity \cite{leutnonrel}) and the
dispersion law reads $\omega \propto |{\bf q}|$.  The existence of a
non-trivial source for the charge density is needed for building up an
$(n_t,n_s)$ effective lagrangian with odd integer values for
$n_t$. Transcribing these principles to the low-energy pion
propagation in an isotropic and homogenous nuclear matter background,
we see that we can expect the `ferromagnetic' dispersion in case we
have a non-vanishing external isovector source, i.e.\ the propagation
in isospin non-symmetric nuclear matter where the presences of the
isovector-density implies that the Weinberg-term (an $\cO{\omega}$
term) is governing the pion evolution.  However, for isospin symmetric
nuclear matter, all external isovector sources are zero such that
Weinberg-type terms are ineffective and the pion propagation -- in the
chiral limit -- has to follow the `antiferromagnetic dispersion' law
of lagrangians of proto-type
\equa{ln20} and \equa{ln02}. Rewriting
\equa{ln20} and \equa{ln02} in
the standard quaternion-formulation, one
gets the following structure \cite{leutnonrel}
\be
 {\cal L}_{\rm eff}^{\rm a.f.m.} =
\frac{ F_1^2}{4} \Tr( \del_0 U \del_0 U^\dagger)
               -\frac{ F_2^2}{4} \Tr( \del_i U \del_i U^\dagger)
    +\cO{\omega^4,\omega^2 |{\bf q}|^2, |{\bf q}|^4}  \
\ee
with two effective coupling constants $F_1$ and $F_2$ such that the
dispersion law corresponds to a massless particle (as there is no
explicit symmetry breaking included) moving with the velocity
$v=F_2/F_1$: $ \omega ({\bf q}) = v\, |{\bf q}| + \cdots$. (Note
$c\!\equiv\!1$.)\ Now, our mean-field lagrangian \equa{lrho} can be
recasted into the following form
\be
 {\cal L}_{\rm eff}^{\rm m.f.}
 &=& \frac{ {f_t^\ast}^2(\rho)}{4} \Tr( \del_0 U \del_0 U^\dagger)
         -\frac{ {f_s^\ast}^2(\rho)}{4} \Tr( \del_i U \del_i U^\dagger)
 \nno \\
&& \mbox{}
  +\frac{f^2_\pi}{4}\, \frac{B^\ast(\rho)}{B}\,
 \Tr(U\chi^\dagger\mbox{+}U^\dagger\chi)
 +\cO{m_\pi^3;\rho^{\nu}|_{\nu>1}} \ .
  \label{lmf_gen}
\ee
Note \equa{lmf_gen} is consistent to $\cO{Q^2}$ (and in the chiral
limit) with the general expression of the leading-order
non-relativistic antiferromagnetic effective lagrangian.  Thus the
in-medium lagrangian for pion propagation in symmetric nuclear matter
is to this order the transcription of the standard Lorentz-invariant
non-linear $\sigma$-model to a spatially rotational-invariant
generalization with two -- in general -- different density-dependent
coefficients. In the same line of thought one would naively expect
that the next corrections correspond to the generalizations of the ten
SU(2) vacuum terms of $\cO{Q^4}$ (see Ref.\cite{gl}) to eighteen terms
-- again with density-dependent coefficients -- which would scale as
$\cO{\omega^4}$, $\cO{\omega^2|{\bf q}|^2}$, $\cO{|{\bf q}|^4}$,
$\cO{\omega^2 m_\pi^2}$, $\cO{|{\bf q}|^2 m_\pi^2}$ or $\cO{m_\pi^4}$,
respectively.  However, the mean-field lagrangian predicts that the
first corrections already appear at order $\cO{m_\pi^3}$ which naively
seems to be in contradiction with the general scheme based on
isospin-symmetric background matter: First, the non-existence of
isospin-symmetry breaking sources together with the chiral symmetry
excludes kinetic lagrangians of odd order in $n_t$ (odd terms in $n_s$
are anyhow excluded because of the spatial rotational symmetry of the
isotropic background). Second, 1-loop corrections to the 2nd order
lagrangian have to be of $\cO{\omega^4}$, $\cO{{\bf q}^4}$ or
$\cO{\omega^2 {\bf q}^2}$ and therefore of $\cO{m_\pi^4}$ (if $\omega$
and $|{\bf q}|$ are counted as $\cO{m_\pi}$).  How can $\cO{m_\pi^3}$
correction terms show up in the general scheme? We know they have to
as they already are present on the simple mean-field level to linear
order in the density. The answer is we need a ${\cal L}^{(3)}$
lagrangian which cannot be purely kinetic and which is not present in
the standard formulation of chiral perturbation theory \cite{gl}.
Fortunately, chiral perturbation theory can be generalized to
incorporate even such terms~\cite{ks}.  One of the preconditions on
standard chiral perturbation is that the parameter $B$ (~defined in
the scalar-pseudoscalar source $\chi =2 B(s +i p)$~) is of order of
the chiral symmetry breaking scale $\Lambda \simeq 1$~GeV. In other
words, as $B$ determines the two-quark condensate, the precondition is
that the magnitude of $\langle \bar q q\rangle$ is large compared with
the four quark condensate such that the GMOR relation holds and that
therefore $m_\pi^2 \propto \hat m = (m_u+m_d)/2$. Then the latter
relation implies that $\hat m$ has to be counted as
$\cO{Q^2}$. However, as we have seen from the model-independent
relation \equa{gmorrho}, the in-medium quark condensate decreases
rather rapidly: Already at nuclear matter densities it drops to about
two-thirds of its vacuum value. Thus, it is not obvious any longer
that the in-medium four-quark condensate can be safely neglected at
$\cO{Q^2}$.  In Ref.\cite{ks} an extension to the GMOR relation is
discussed; i.e.\ terms of order $\langle 0| \bar q q \bar q q
|0\rangle $ are taken into account so that the pion mass square is
parameterized as $m_\pi^2$ = $\hat m B_0\mbox{+}{\hat m}^2 A_0$, where
$A_0$ is related to the four-quark condensate. In the vacuum (~and for
SU(2)~) $B_0/2 A_0 \gg \hat m$, so the second term is of no
importance.  {\em But} if one now goes to {\em finite density}, the
quark condensate $-\langle \bar q q \rangle_\rho$ decreases in value.
Thus the effect of the four-quark condensate doesn't need to be small
any longer compared with the two-quark condensate.  This has
consequences for an in-medium effective chiral lagrangian: In the
vacuum, and for SU(2), it is not necessary to include in the
leading-order effective lagrangian the four-quark condensate directly
(as in Ref.\cite{ks} ) as the two-quark condensate is large and
doesn't change. But in the medium the two-quark condensate changes and
decreases in value (it can even become vanishingly small), such that
one has to offer to the in-medium effective lagrangian the possibility
that the four-quark condensate determines the effective pion mass,
too.  This demands {\bf a)} the generalization of the ${\cal L}^{(2)}$
terms (to include three additional contributions) and {\bf b)} the
inclusion of eleven distinct ${\cal L}^{(3)}$ terms as first
correction, see Ref.\cite{ks}; both changes will modify the r.h.s.\ of
the GMOR-like relation.  In general, the expansion of the vacuum
effective lagrangian
\be
  {\cal L}_{\rm eff}^{\rm vac} =\!\! \sum_{\nu=2,4,6,\dots}
 \sum_{{\scriptscriptstyle
{\renewcommand{\arraystretch}{0.6}\begin{array}[t]{c}
                          {\scriptscriptstyle   k,l\geq 0}\\
                            {\scriptscriptstyle 2(k+l)=\nu}
                            \end{array}}}}
   {\cal L}^{(\nu)}_{2k;l} \quad {\rm with}\ \,  {\cal L}^{(\nu)}_{2k;l}
 \sim  Q^{2k} ( B_0 {\hat m} )^l \nno
\ee \vskip -2mm\noindent
should be replaced in isospin symmetric, isotropic and homogenous
matter by the expansion
\be
{\widetilde {\cal L}}_{\rm eff}(\rho)=\!\!\!
\sum_{\nu=2,3,4,5,\dots} \! \!\!
\sum_{{\scriptscriptstyle {\renewcommand{\arraystretch}{0.6}
            \begin{array}[t]{c}
                            {\scriptscriptstyle j,k,l,n\geq 0}\\
                             {\scriptscriptstyle 2(j+k+l)+n=\nu}
                            \end{array}}}} \!\!\!\!\!\!\!
 {\widetilde {\cal L}}^{(\nu)}_{2j,2k;l,n}(\rho)
\quad {\rm with}\ \,   {\widetilde{\cal L}}^{(\nu)}_{2j,2k;l,n}(\rho)
 \sim  \omega^{2j}  |{\bf q}|^{2k} ( B^\ast( \rho )\hat m)^l
    {\hat m}^n   . \nno
\ee \vskip -2mm\noindent
Thus {\bf a)} the coefficients of the various lagrangians are
density-dependent, {\bf b)} Lorentz-invariance is broken down to just
spatial rotational invariance such that $Q^2$-dependent contributions
split into $\omega^2$- and $\bf q^2$-dependent ones, and {\bf c)} also
{\em totally new} terms appear which have no counter parts in the
vacuum sector of standard chiral perturbation theory.
One might speculate that the occurrence of an $\cO{Q^3}$ correction in
the mean-field approximation is already a signal or precursor of the
change in the relation between ${m^\ast_\pi}^2(\rho)$ and the current
quark mass matrix ${\cal M}$ from that given by the vacuum GMOR
relation. This phenomenon has eventually to take place as with
increasing density the effective pion comes closer and closer to
chiral restoration. In the framework of a linear $\sigma$-model its
mass has to approach the effective mass of the $\sigma$ (the chiral
partner of the pion).  The latter mass scales as ${\cal O}(\hat m)$
and not as ${\cal O}(\sqrt{\hat m})$. Stated differently, with
increasing density the pion should behave more and more like a
`normal' meson (like the $\rho$-meson or $\omega$) as chiral symmetry
becomes more and more restored. If the pion behaves more `normal' in
this respect, its effective mass has also to scale more `normally'.

In summary, additional ${\cal L}^{(3)}(\rho)$ terms are needed in
nuclear matter to incorporate the changed scaling behavior of the pion
and the fact that the in-medium quark condensate decreases in
value. This effect is already present at  linear order in density:
In Ref.\cite{bkm} finite ${\cal O}(m_\pi^3)$ loop terms were found
which `renormalize' the isospin even scattering length and which lead
to $\cO{Q^3}$ corrections in the mean-field approximation.  The new
${\cal L}^{(3)}(\rho)$ terms and the additional ${\cal L}^{(2)}(\rho)$
terms of the then necessary generalization of chiral perturbation
theory (see Ref.\cite{ks}) will make it rather unlikely that the GMOR
relation prevails at high nuclear densities.  In fact, the vanishing
of the in-medium quark condensate which is linked to the vanishing of
the PCAC-source coupling in \equa{lgl} and the vanishing of the
residuum of the in-medium propagator \equa{gssprop} should be rather
interpreted as a signal that the standard chiral perturbation theory
is pushed beyond its limits of applicability. The theory {\em has} to
be generalized~\footnote{The theory must be generalized even further,
in case an in-matter resonance {\em pole} becomes comparable with the pion
effective mass. Note, however, because of the S-wave nature of the
pion propagation additional {\em cuts} are not important, at least at low
densities.}.  Unfortunately, all the coefficients which enter the
generalized in-medium chiral perturbation theory are density-dependent
with a priori unknown free functional forms, furthermore the
generalized theory incorporates terms which do not have any vacuum
analogs.  So, the only chance of getting constraints on these free
density-dependent coefficients is from the rather scarce astrophysical
input and perhaps also from heavy-ion-scattering data which, however,
are ``polluted'' by temperature dependences.  In practice, the virtues
of the generalized in-medium effective lagrangians are rather the
constraints which they pose on hadronic models applied to the pion (or
kaon) propagation in the nuclear medium, since chiral perturbation
theory gives intercorrelations between data and therefore
model-independent results which specific to-be-tested models might
either follow or not.

\vskip 2.5mm
\ni {\bf Acknowledgements}\\[1mm]
We are grateful to F. Beck,
H. Bijnens, G. E. Brown, N. Kaiser, M.~Kirchbach, M.~Lutz,
A.~Manohar, U.-G. Mei{\ss}ner,
M. Rho, W. Weise and I. Zahed for
discussions.

\renewcommand{\baselinestretch}{1.0}

{\small
\begin{thebibliography}{99}
%
\bibitem{kapnel}
D.B. Kaplan and A.E. Nelson, Phys. Lett. {\bf B175} (1986) 57
%
\bibitem{bkr}
G. E. Brown, V. Koch and M. Rho, Nucl. Phys. {\bf A535} (1991) 701
%
\bibitem{dee}
J. Delorme, M. Ericson and T.E.O. Ericson, Phys. Lett. {\bf B291} (1992) 379
%
\bibitem{blrt}
G. E. Brown, C.-H. Lee, M. Rho and V. Thorsson,
Nucl. Phys. {\bf A567} (1993) 937
%
\bibitem{sc1}
H. Yabu, S. Nakamura, F. Myhrer and K. Kubodera,
Phys. Lett. {\bf B315} (1993) 17; H. Yabu, S. Nakamura
and K. Kubodera, Phys. Lett. {\bf B317} (1993) 269;
H. Yabu, F. Myhrer and K. Kubodera, Phys. Rev. {\bf D50} (1994) 3549
%
\bibitem{we} V. Thorsson and A. Wirzba,
{\em S-wave Meson-Nucleon Interactions
and the Meson Mass in Nuclear Matter from Chiral Effective
Lagrangians}, NORDITA-95/7 N, {\tt nucl-th/9502003}
%
\bibitem{ccwz}  S. Coleman, J. Wess and B. Zumino, Phys. Rev.
{\bf 177} (1969) 2239
%
\bibitem{gl}
J. Gasser and H. Leutwyler, Ann. Phys. (N.Y.) {\bf 158} (1984) 142
%
\bibitem{leutnonrel}
H. Leutwyler, Phys. Rev {\bf D49} (1994) 3033
%
\bibitem{ks}
M. Knecht and J. Stern, IPNO-TH-94-53, to be published in
the second edition of the DAPHNE physics handbook, Eds. L. Maini, G. Pancheri
and N. Paver, {\tt hep-ph/9411253}
%
\bibitem{gss}
J. Gasser, M.E. Sainio and A. \u{S}varc, Nucl. Phys. {\bf B307} (1988) 779
%
\bibitem{weinberg} S. Weinberg, Physica {\bf A96} (1979), 327; Phys. Lett.
{\bf B251} (1990) 288; Nucl. Phys. {\bf B363} (1991) 3
%
\bibitem{jenman}
E. Jenkins and A. Manohar, Phys. Lett. {\bf B255} (1991) 558
%
\bibitem{bkkm}
V. Bernard, N. Kaiser, J. Kambor and U.-G. Mei{\ss}ner,
Nucl. Phys.  {\bf B388} (1992) 315.
%
\bibitem{bkm}
V. Bernard, N. Kaiser and U.-G. Mei{\ss}ner, Phys. Lett. {\bf B309}
(1993) 421
%
\bibitem{gls}
J. Gasser, H. Leutwyler and M. E. Sainio,
Phys. Lett. {\bf B253} (1991) 252.
%
\bibitem{koch}
R. Koch, Nucl. Phys. {\bf A448} (1986) 707
%
\bibitem{polerefs}
G. Baym and E. Flowers, Nucl. Phys. {\bf A222}, (1974) 29;
A. E. Nelson and D. B. Kaplan, Phys. Lett. {\bf B192} (1987) 193;
T. Muto and T. Tatsumi, Phys. Lett. {\bf B283} (1992) 165;
V. Thorsson, M. Prakash and J.M. Lattimer,
Nucl. Phys. {\bf A572} (1994) 693; {\bf A574} (1994) 851
%
\bibitem{gmor}
M. Gell-Mann, R. J. Oakes and B. Renner,
Phys. Rev. {\bf 175} (1968) 2195.
%
\bibitem{kirch}
M. Kirchbach and D. O. Riska, Nucl. Phys. {\bf A578} (1994) 511
%
\bibitem{birse}
E. G. Drukarev and E. M. Levin, Nucl. Phys. {\bf A511} (1988) 697;
T. D. Cohen, R. J. Furnstahl and D. K. Griegel, Phys. Rev. Lett.
{\bf 67} (1991); Phys. Rev {\bf C45} (1992) 1881;
M. C. Birse, J. Phys. G. {\bf 20} (1994) 1537
%
\bibitem{gmorrho}
V. Bernard and U.-G. Mei{\ss}ner, Nucl. Phys. {\bf A489} (1988) 647;
M. Lutz, A. Steiner and W. Weise, Nucl. Phys. {\bf A542} (1992) 521;
{\bf A574} (1994) 755
%
\end{thebibliography}
}
\end{document}